\documentclass[%
 reprint,
 superscriptaddress,
 amsmath,amssymb,
 aps,
prl,
]{revtex4-1}

\usepackage[usenames]{color}
\usepackage{graphicx}
\usepackage{dcolumn}
\usepackage{bm}

\begin{document}

\preprint{APS/123-QED}
\title{Single Spin Asymmetries in Charged Pion Production from
Semi-Inclusive Deep Inelastic Scattering on 
a Transversely Polarized $^3$He Target at ${\bf Q^2=1.4-2.7}$ GeV$^2$}


\author{X.~Qian} \email[Corresponding author: ]{xqian@caltech.edu}
\affiliation{Duke University, Durham, NC 27708}
\affiliation{Kellogg Radiation Laboratory, California Institute 
  of Technology, Pasadena, CA 91125}
\author{K.~Allada}
\affiliation{University of Kentucky, Lexington, KY 40506}
\author{C.~Dutta}
\affiliation{University of Kentucky, Lexington, KY 40506}
\author{J.~Huang}
\affiliation{Massachusetts Institute of Technology, Cambridge, MA 02139}
\author{J.~Katich}
\affiliation{College of William and Mary, Williamsburg, VA 23187}
\author{Y.~Wang}
\affiliation{University of Illinois at Urbana-Champaign, Urbana, IL 61801}
\author{Y.~Zhang}
\affiliation{Lanzhou University, Lanzhou 730000, Gansu, People's Republic of China}
\author{K.~Aniol}
\affiliation{California State University, Los Angeles, Los Angeles, CA 90032}
\author{J.R.M.~Annand}
\affiliation{University of Glasgow, Glasgow G12 8QQ, Scotland, United Kingdom}
\author{T.~Averett}
\affiliation{College of William and Mary, Williamsburg, VA 23187}
% ...
\collaboration{The Jefferson Lab Hall A Collaboration}
\noaffiliation

\date{\today}

\begin{abstract}
We report the first measurement of target single spin asymmetries
in the semi-inclusive
$^3\mbox{He}(e,e'\pi^\pm)X$ reaction on a
transversely polarized target. The experiment, conducted
at Jefferson Lab using a 5.9 GeV electron beam, covers a
range of 0.16 $< x <$ 0.35 with 1.4 $<Q^2<$ 2.7 GeV$^2$. The
Collins and Sivers moments were extracted from the azimuthal angular dependence
of the measured asymmetries. The $\pi^\pm$ Collins moments for
$^3$He are consistent with zero, except for the $\pi^+$ moment at
$x=0.35$, which deviates from zero by 2.3$\sigma$. While the
$\pi^-$ Sivers moments are consistent with zero, the $\pi^+$ Sivers
moments favor negative values. The neutron results were extracted
using the nucleon effective polarization and measured cross section
ratios of proton to $^3$He,
and are largely consistent with the predictions of
phenomenological fits and quark model calculations. 



\end{abstract}
\pacs{Valid PACS appear here}

\maketitle
%
%
%
%
%
%
%
%
%
%
%
%
%
%
%

\vspace{5.0cm}


High-energy lepton-nucleon scattering is a powerful tool to study the
partonic structure of the nucleon. While detailed studies of inclusive
deep inelastic scattering (DIS) have revealed a great deal of
information about the unpolarized ($f_1^q$) and polarized ($g_1^q$)
parton distribution functions (PDFs) describing the longitudinal
momentum and helicity of quarks in the nucleon, understanding of 
the  nucleon's spin structure is far from being complete~\cite{spin_review}. 
In particular, the experimental study of quark transverse spin phenomena has just 
begun~\cite{hermes-trans,hermes-trans-new,compass-new,compass-proton}.  
Recent reviews can be found in Ref.~\cite{ssa_review, ssa_review1}.
These progresses also point to an important role for quark/gluon 
orbital angular motion in the nucleon's spin structure. Semi-inclusive DIS 
(SIDIS), in which a hadron from the fragmentation 
of the struck quark is detected
in coincidence with the scattered lepton, provides access to transverse-momentum-dependent 
parton distributions (TMDs)~\cite{Mulder2,mulders,Bacche}, 
which describe the quark structure of the nucleon in all three 
dimensions of momentum space. The ability of SIDIS reactions to access 
partonic transverse spin and momentum~\cite{TMD_fact,HallC_SIDIS,CLAS,hermes-trans,hermes-new,
compass-proton,compass-new} relevant to the kinematics of this work provides 
a unique opportunity for the study of orbital angular momentum (OAM).


All eight leading-twist TMDs are accessible in SIDIS~\cite{Bacche}. 
The angular dependence of the target spin-dependent asymmetry $A$ in the 
scattering of an unpolarized lepton beam by a transversely polarized 
target is:
\begin{eqnarray}\label{eq:asy}
A(\phi_h,\phi_S) &=& \frac{1}{P}\frac{Y_{\phi_h,\phi_S}-Y_{\phi_h,\phi_S +
    \pi}}{Y_{\phi_h,\phi_S} + Y_{\phi_h,\phi_S + \pi}} \nonumber \\
&\approx&  A_{C} \sin(\phi_h + \phi_S) + A_{S}\sin(\phi_h -
    \phi_S),  
\end{eqnarray}
where $P$ is the target polarization, $\phi_h$ and $\phi_S$ are the
azimuthal angles of the hadron momentum and the target spin relative
to the lepton scattering plane as defined in the Trento 
convention \cite{trento}, 
$Y$ is the  normalized yield, and $A_C$ ($A_S$) is the Collins (Sivers)
moment.




The Collins moment probes the convolution of the chiral-odd
quark transversity distribution $h_{1}^q$~\cite{transversity2} 
and the chiral-odd Collins fragmentation function (FF)~\cite{collins3}. 
$h_{1}^q$ describes 
the transverse polarization of quarks in a transversely 
polarized nucleon. 
Because the gluon transversity vanishes, 
quark transversity is valence-like~\cite{nog_trans}. The 
lowest moment of transversity, the tensor charge, provides a test 
of lattice QCD predictions~\cite{lattice}. 
Transversity is further constrained by Soffer's inequality~\cite{soffer}, 
$\left|h_{1}^q\right|\le \frac{1}{2}\left(f_1^q + g_1^q\right)$,
which holds under next-to-leading-order QCD 
evolution~\cite{Kumano:1997qp,Hayashigaki:1997dn,werner}. 
However, a possible violation of Soffer's bound has been 
suggested~\cite{doubt}.




The Sivers moment probes the convolution of the naive T-odd quark
Sivers function $f^{\perp}_{1T}$~\cite{sivers1} and the unpolarized FF. 
$f^{\perp}_{1T}$ represents a correlation between
the nucleon spin and the quark transverse momentum, and it corresponds
to the imaginary part of the interference between light-cone wave function 
components differing by one unit of OAM~\cite{brodsky,OAM}. 
The Sivers function was originally thought to vanish 
since it is odd under naive time-reversal transformations~\cite{collins3}.
A nonzero $f_{1T}^{\perp}$ was later shown to be allowed due to QCD
final state interactions (FSI) between the outgoing quark 
and the target remnant~\cite{brodsky}. It was further demonstrated through 
gauge invariance that the same Sivers function, which originates from 
a gauge link, would appear in both SIDIS and Drell-Yan single spin asymmetries 
(SSAs) but with an opposite sign~\cite{Collins_siver,brodsky_siver}.


The HERMES collaboration carried out the first SSA measurement in
SIDIS on a transversely polarized proton target using $e^\pm$ beams
\cite{hermes-trans} at $Q^2=1.3-6.2$ GeV$^2$. The COMPASS collaboration performed SIDIS 
measurements with a muon beam on transversely polarized deuteron
\cite{compass-new} and proton \cite{compass-proton} targets at $Q^2=1.3-20.2$ GeV$^2$. Large
Collins moments were observed for both $\pi^+$ and $\pi^-$ from the proton, 
but with opposite sign, indicating that the ``unfavored''
Collins FF could be as large as the ``favored'' one~\cite{collins3}. 
This finding is consistent with the measured asymmetry of
inclusive hadron pair production in $e^+e^-$ annihilation from 
BELLE~\cite{belle}, which directly accessed the product of Collins FFs. 
The deuteron Collins asymmetries for $\pi^+$ and $\pi^-$ are
 consistent with zero, but with relatively large uncertainties for $x$ $>$ 0.1, 
which suggests a cancellation between proton and neutron. 


While both the HERMES and COMPASS proton data show significantly positive 
$\pi^+$ Sivers moments, a possible inconsistency exists 
between the data sets~\cite{ans_2010a}. On the other hand,
the proton  $\pi^-$ Sivers moments from both HERMES~\cite{hermes-new} and 
COMPASS~\cite{compass-proton} are consistent with zero, along with the
 COMPASS deuteron $\pi^+$ and $\pi^-$ Sivers moments.  \textcolor{black}{These results could 
reflect pronounced flavor dependence of the Sivers functions, as indicated
by a phenomenological fit~\cite{ans_2010a} of these data.}
\textcolor{black}{To shed new light on the flavor structures of the transversity and
Sivers functions, it is important to extend the SSA SIDIS measurement to
a neutron target, which is more sensitive to the nucleon's $d$ quark contribution.}
Since there is no stable free neutron target, polarized $^3$He is 
commonly used as an effective polarized neutron target \cite{he3_2}. 
The $^3$He nucleus, in which the nuclear spin resides predominantly with
the neutron, is uniquely advantageous in the extraction of
neutron spin information compared to the deuteron (p+n).

In this letter, we present the results of SSA measurements in SIDIS on a 
transversely polarized $^3$He target, performed in Jefferson Lab
(JLab) Hall A from 2008/11 to 2009/02. 
The electron beam energy was 5.9 GeV with an average current of 12 $\mu A$. 
Scattered electrons with momenta from 0.6--2.5 GeV were detected in the BigBite
spectrometer at a central angle of 30$^\circ$ on the beam
right. Coincident charged hadrons were detected in the High Resolution
Spectrometer (HRS) \cite{halla_nim} at a central angle of 16$^\circ$ on beam left
and a central momentum of 2.35 GeV. 
Unpolarized beam was achieved by summing the two beam helicity
states. The residual beam charge asymmetry was smaller than 100 ppm per 1-hour run.

The 40 cm long polarized $^3$He~\cite{halla_nim} 
cell was filled at room temperature with $\sim$8 atms
of $^3$He and $\sim$0.13 atms of N$_2$ to reduce depolarization 
effects. 
The $^3$He nuclei were polarized by Spin Exchange Optical Pumping 
of a Rb-K mixture. The polarization was monitored 
by Nuclear Magnetic Resonance (NMR) measurements every 20 minutes as 
the target spin was automatically flipped through Adiabatic Fast 
Passage. The NMR measurements were calibrated using the known
water NMR signal and cross-checked using the Electron Paramagnetic
Resonance method. The average polarization was 55.4$\pm$2.8\%. 
Three pairs of mutually orthogonal Helmholtz coils were used to orient the target
polarization vertically and horizontally (determined to 
better than 0.5$^\circ$ using a compass) in the plane transverse to
the beam direction in order to maximize the $\phi_S$ coverage.
The holding magnetic field ($\sim$25 G) remained fixed during spin flips. 


\begin{table*}[ht]
\centering
\begin{tabular}{|c|c|c|c|c|c|c|c|c|c|c|}\hline
$x$ & $Q^2$ & $y$ & $z$ & $P_{h\perp}$ & W & W$^{\prime}$ & $f^{\pi^+}_{pair}$ &$f^{\pi^-}_{pair}$  & $1-f^{\pi^+}_{p}$ & $1-f^{\pi^-}_{p}$\\
& GeV$^2$ & & & GeV & GeV& GeV & & & &\\\hline
 0.156~ & ~1.38~ & ~0.81~ & ~0.50~ & ~0.435~ &~2.91~&~2.07~&22.0$\pm$4.4\% & 24.0$\pm$4.8\% & ~$0.212\pm0.032~(0.027)$~&~$0.348\pm0.032~(0.022)$~\\
 0.206~ & ~1.76~ & ~0.78~ & ~0.52~ & ~0.38~ &~2.77~&~1.97~ &8.0$\pm$2.0\% & 14.0$\pm$2.0\% & ~$0.144\pm0.031~(0.029)$~&~$0.205\pm0.037~(0.027)$~\\
 0.265~ & ~2.16~ & ~0.75~ & ~0.54~ & ~0.32~ &~2.63~&~1.84~ &2.5$\pm$0.9\% & 5.0$\pm$1.8\% & ~$0.171\pm0.029~(0.028)$~&~$0.287\pm0.036~(0.024)$~\\
 0.349~ & ~2.68~ & ~0.70~ & ~0.58~ & ~0.24~ &~2.43~&~1.68~ &1.0$\pm$0.5\% & 2.0$\pm$1.0\% & ~$0.107\pm0.026~(0.030)$~&~$0.220\pm0.032~(0.026)$~\\\hline
\end{tabular}
\centering
  \caption{Central kinematics for the four $x$ bins. The fractional
    $e^-$ energy loss $y$, the hadron energy fraction $z$ with respect of electron energy transfer
    and the transverse momentum $P_{h\perp}$ are all defined following
    the notation of Ref.~\cite{Bacche}.
   The pair production background $f^{\pi^\pm}_{pair}$  and the 
   proton dilution $1-f^{\pi^\pm}_{p}$ are shown with their total
   experimental systematic uncertainties. \textcolor{black}{The 
     numbers in parentheses represent the model uncertainties
     corresponding to unpolarized FSI effects.}}
\label{table:kine}
\end{table*}

The BigBite spectrometer consists 
of a large-opening dipole magnet  in front of a detector
stack including three sets of multi-wire drift chambers 
for charged-particle tracking, a  lead-glass calorimeter divided 
into preshower/shower sections for electron identification and a 
scintillator plane between the preshower and shower for timing. 
In this experiment, BigBite was positioned to subtend a solid
angle of $\sim$64 msr for a 40 cm target. The large out-of-plane angle
acceptance of BigBite ($\pm$240 mrad) was essential in maximizing the $\phi_h$
coverage of the experiment, given the small ($\sim$6 msr) solid angle
acceptance of the hadron arm. 
The transport matrix of the BigBite magnet was calibrated using
a multi-foil carbon target, a sieve slit collimator and $^1$H$(e,e')p$
elastic scattering at incident energies of 1.2 and 2.4 GeV. The achieved angular
and momentum resolutions were better than 10 mrad and 1\%, respectively. Clean $e^-$ identification was achieved using
cuts on the preshower energy $E_{ps}$ and the 
ratio $E/p$ of the total shower energy to the momentum from optics 
reconstruction. The $\pi^-$ contamination was determined from analysis of the
$E_{ps}$ spectrum to be less than 2\%, consistent with 
GEANT3 simulations. 

The HRS detector package was configured for hadron detection \cite{halla_nim}. 
A 10$^4$:1 $e^-$ rejection factor was achieved using a light gas \v{C}erenkov 
and a lead glass calorimeter, resulting in a negligible $e^\pm$ contamination. Coincidence timing provided more than 15$\sigma$ 
pion-proton separation. A 10:1 $K^\pm$ rejection was achieved using the 
aerogel \v{C}erenkov detector, leaving less than 1\%
contamination. The $\pm5\%$ HRS momentum acceptance limited the hadron
energy fraction $z$ to about $0.5$ (see Table \ref{table:kine}).
SIDIS events were selected using cuts on the four-momentum
transfer squared $Q^2 > $ 1 GeV$^2$, the hadronic final-state
invariant mass W $>$ 2.3 GeV, and the mass of undetected
final-state particles W$^{\prime}$ $>$ 1.6 GeV, assuming scattering on a nucleon. 
The total number of accepted SIDIS events are 254k and 194k for $\pi^+$ and $\pi^-$, respectively.
The data were divided into four
bins in the Bjorken scaling variable $x$. The central kinematics of
the four bins \textcolor{black}{after radiative corrections} are presented in Table \ref{table:kine}. 
SIDIS yields were obtained by normalizing the number of identified
SIDIS events by the accumulated beam charge and the data acquisition
live time. The data were divided into $\sim$2850 pairs of measurements in opposite target spin
states to extract the raw asymmetries. The false asymmetry due to luminosity fluctuations was  
confirmed to be less than 4$\times10^{-4}$ by 
measurements of the SSA in
inclusive $(e,e')$ scattering with transverse target polarization oriented 
horizontally, which vanishes due to parity conservation.
The raw Collins/Sivers moments
were obtained by fitting the asymmetries in 2-D ($\phi_h,
\phi_S$) bins according to Eq.\eqref{eq:asy}. 
This procedure was confirmed by an unbinned maximum-likelihood method. 
The $^3$He  moments were obtained after correcting  the directly measured 
N$_2$ dilution ($\sim$10\% contribution). 


\begin{figure}[]
\includegraphics[width=90mm]{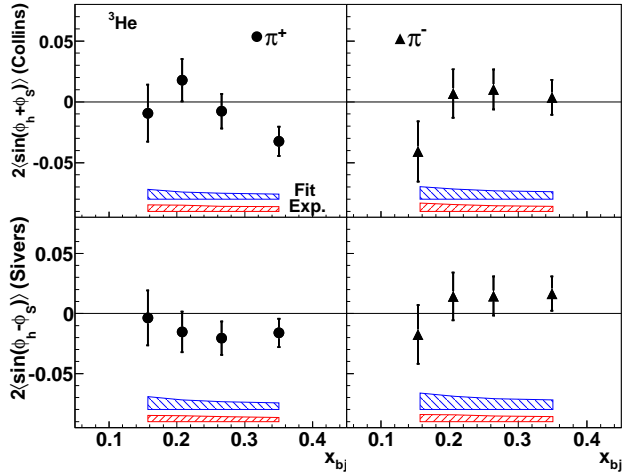}
\caption{\label{fig:3he} (Color online) The extracted Collins/Sivers moments on $^3$He are 
  shown together with uncertainty bands (see text) for both $\pi^+$ and 
	$\pi^-$ electro-production.  
} 
\end{figure}

The dominant background in the SIDIS electron sample comes from
$e^+/e^-$ pair production. 
This background (listed in Table~\ref{table:kine}) was directly measured 
by reversing the polarity of the BigBite magnet to detect $e^+$ in 
identical conditions as $e^-$. The contamination
was treated as a dilution effect in the analysis, as the measured asymmetries
were consistent with zero for $e^+$-$\pi$ coincidence events, 
which mirror the pair-produced $e^-$-$\pi$ events.
Additional experimental uncertainties in the extracted
$^3$He Collins/Sivers moments include: 
1) $K^\pm$ contamination in the $\pi^\pm$ sample, 
2) bin-centering, resolution and radiative effects estimated
using simulations, 
3) the effect of the target
collimator, estimated by varying the scattering vertex cut, 
4) target density fluctuations, and 
5) the false asymmetry due to yield drift caused by radiation damage
to the BigBite preshower calorimeter. 
The quadrature sum of all above contributions 
is below 25\% of the statistical uncertainty in each $x$ bin.

In addition, there
are fitting systematic uncertainties resulting from the neglect of 
other $\phi_h$- and $\phi_S$-dependent terms, 
such as $2\langle\sin (3\phi_h-\phi_S)\rangle$, higher-twist terms including $2\langle\sin \phi_S\rangle$ and
$2\langle\sin\left(2\phi_h-\phi_S\right)\rangle$, azimuthal
modulations of the unpolarized cross section including the 
Cahn ($2\langle\cos \phi_h\rangle$) and
Boer-Mulders (2$\langle\cos (2\phi_h)\rangle$) 
effects~\cite{Bacche}, and leakage from the longitudinal SSA (A$_{UL}$)
due to the small longitudinal component of the target polarization.  
The effects of these terms were 
estimated by varying each term within an allowed range derived
from the HERMES data~\cite{hermes_thesis1,hermes_thesis2}, assuming the magnitude of
each term for the neutron is similar to that of the
proton. The $2\langle\sin \phi_S\rangle$ term 
gives the largest effect, followed by the $2\langle\sin (3\phi_h-\phi_S)\rangle$  and 
$2\langle\sin\left(2\phi_h-\phi_S\right)\rangle$ terms.

A Monte Carlo simulation of the experiment was adapted from the package SIMC used
in the analysis of SIDIS cross section measurements on $^1$H and $^2$H
from JLab Hall C
\cite{HallC_SIDIS} to include models of our target and spectrometers. 
SIMC was used to estimate the combined effects of acceptance, resolution
and radiative corrections on the extraction of the Collins and Sivers moments, and these effects were included
in the experimental systematic uncertainties. Additionally, the
contamination in identified SIDIS events from decays of diffractively
produced $\rho$ mesons was estimated to range from 3-5\% (5-10\%) for
$\pi^+$ ($\pi^-$) by PYTHIA6.4 \cite{pythia}. Consistent with the HERMES analysis, 
no corrections for this background have been applied to our results.
\textcolor{black}{The contamination from radiative tails of exclusive 
electroproduction, estimated by normalizing the MC spectrum to the data in the 
low-$W$ region, was found to be less than 3\%.}
The extracted $^3$He Collins $A_{C}\equiv 2\langle\sin(\phi_h+\phi_S)\rangle$ and 
Sivers $A_{S}\equiv 2\langle\sin(\phi_h-\phi_S)\rangle$ moments are shown in Fig.~\ref{fig:3he}
and tabulated in Table.~\ref{table:results}. 
The error bars represent statistical uncertainties only. 
The experimental systematic uncertainties combined in quadrature are
shown as the band labeled ``Exp.''. The combined extraction model 
uncertainties due to neglecting other allowed terms are shown 
as the band labeled ``Fit''. \textcolor{black}{The extracted $^3$He Collins and
  Sivers moments are all below 5\%. }
The Collins moments are mostly consistent with zero, except 
the $\pi^+$ Collins moment at $x$=0.35, which deviates 
from zero by 2.3$\sigma$ after combining the statistical 
and systematic uncertainties in quadrature. 
The $\pi^+$ Sivers moments favor
negative values, and the $\pi^-$ Sivers moments are 
consistent with zero.


\textcolor{black}{To extract the neutron Collins/Sivers SSAs ($A_{n}^{C/S}$)
from the measured $^3$He moments ($A_{\rm ^{3}He}^{C/S}$), we 
used,
\begin{equation}
A_{\rm ^3He}^{C/S} = P_n \cdot (1-f_p) \cdot A_{n}^{C/S}  + P_p f_p
\cdot A_{p}^{C/S} \label{eqn:effpol},
\end{equation}
which was shown to be valid in a calculation by Scopetta \cite{Scopetta} 
including initial-state nuclear effects.
Here,  $P_n=0.86^{+0.036}_{-0.02}$ ($P_p=-0.028^{+0.009}_{-0.004}$) is the neutron
(proton) effective polarization ~\cite{Zheng_PRC}.
The proton dilution $f_{p} = \frac{2\sigma_p}{\sigma_{\rm ^3He}}$ 
of $^3$He} was measured by comparing the yields of unpolarized 
hydrogen and $^3$He targets in the 
SIDIS kinematics. An additional model uncertainty from spin-independent FSI 
was estimated using pion multiplicity data~\cite{RM} and a Lund
string model-based calculation of the pion absorption
probability~\cite{Lund}. 
An upper limit of 3.5\% on the size of the
FSI effect was used to estimate
the uncertainty in $f_p$, shown in Table \ref{table:kine}, and included
in the ``Fit'' systematic uncertainty. The neutron SSAs due 
to spin-dependent FSI were estimated to be well below 1\% across 
the entire $x$ range with a simple Glauber rescattering model. 

\begin{figure}[]
\includegraphics[width=90mm]{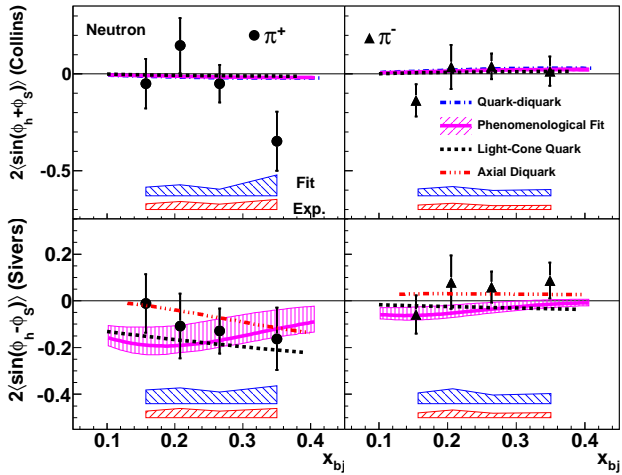}
\caption{\label{fig:neutron} (Color online) The extracted neutron Collins and 
 Sivers moments with uncertainty bands for both $\pi^+$ and $\pi^-$
 electro-production. See text for details.}  
\end{figure}

The resulting neutron Collins/Sivers moments calculated using
Eq.~\eqref{eqn:effpol}, with $f_p$ from our data and proton Collins/Sivers moments from
Refs. \cite{anselmino3,Anselmino:2007fs,ans_2008a}, 
are shown in Fig.~\ref{fig:neutron} and tabulated in Table.~\ref{table:results}. Corrections from the proton Collins/Sivers 
moments are less than 0.012.
%
Our Collins moments are compared with the phenomenological fit~\cite{Anselmino:2007fs}, a light-cone 
quark model calculation \cite{LCQM_C1,LCQM_C2} and quark-diquark model
\cite{CQM_C1,CQM_C2} calculations. The phenomenological fit and the model
calculations, which assume Soffer's bound~\cite{soffer}, predict rather small 
Collins asymmetries which are mostly consistent with our data. 
However, 
the $\pi^+$ Collins moment at $x=0.34$ is suggestive of a noticeably more 
negative value at the $2\sigma$ level. 
\textcolor{black}{Our data favor negative $\pi^+$ Sivers moments, while the $\pi^-$
moments are close to zero. Such behavior 
independently 
supports a negative $d$ quark Sivers function within the parton model picture, 
which has been suggested by 
predictions of the phenomenological fit \cite{anselmino3,ans_2008a}
to HERMES and COMPASS data, a light-cone quark model calculation \cite{LCQM_S1,LCQM_S2},
and an axial diquark model calculation\cite{ADQ}. }

In summary, we have reported the first measurement of the 
SSA in charged pion electroproduction on a transversely polarized
$^3$He target in the DIS region. \textcolor{black}{Our data provide 
the best current measurement of the neutron Sivers moments in 
the valence region ($x>0.1$),
and the best neutron Collins moments for $x>0.2$, 
which will further improve the extraction of
$d$ quark distributions in these regions. }
This experiment has demonstrated the power of
polarized $^3$He as an effective polarized neutron target, and has
laid the foundation for future high-precision measurements of
TMDs with a large acceptance detector SoLID 
following the JLab 12 GeV upgrade~\cite{solid_white} and at an 
electron-ion collider \cite{eic_work}. These future SIDIS data 
taken over a broad range of $Q^2$ will also allow an accurate 
determination of higher twist contribution~\cite{leader,BB}.


We acknowledge the outstanding support of the JLab Hall
A technical staff and the Accelerator Division in accomplishing
this experiment. This work was supported in part
by the U. S. National Science Foundation, and by DOE
contract number DE-AC05-06OR23177, under
which the Jefferson Science Associates (JSA) operates 
the Thomas Jefferson National Accelerator Facility.

\begin{table*}[ht]
\centering
\begin{tabular}{|c|c|c|c|c|c|}\hline
& $x_{bj}$ & Collins Moment  & Collins Moment & Sivers Moment & Sivers Moment \\
& & $\pi^+$ & $\pi^-$ & $\pi^+$ & $\pi^-$ \\\hline
$^3$He & 0.156 & -0.009$\pm$0.023$\pm$0.005 (0.008) & -0.041$\pm$0.025$\pm$0.007 (0.010) & -0.004$\pm$0.023$\pm$0.005 (0.011) & -0.017$\pm$0.025$\pm$0.006 (0.014) \\
$^3$He & 0.206 & 0.018$\pm$0.017$\pm$0.005 (0.006) & 0.007$\pm$0.020$\pm$0.006 (0.008) & -0.015$\pm$0.017$\pm$0.005 (0.008) & 0.014$\pm$0.020$\pm$0.006 (0.011) \\
$^3$He & 0.265 & -0.008$\pm$0.014$\pm$0.004 (0.005) & 0.010$\pm$0.016$\pm$0.005 (0.007) & -0.021$\pm$0.014$\pm$0.004 (0.006) & 0.015$\pm$0.016$\pm$0.005 (0.009) \\
$^3$He & 0.349 & -0.033$\pm$0.012$\pm$0.004 (0.004) & 0.004$\pm$0.014$\pm$0.004 (0.006) & -0.016$\pm$0.012$\pm$0.003 (0.006) & 0.017$\pm$0.014$\pm$0.004 (0.008) \\\hline
n & 0.156 & -0.050$\pm$0.128$\pm$0.029 (0.044) & -0.137$\pm$0.084$\pm$0.023 (0.036) & -0.011$\pm$0.125$\pm$0.028 (0.059) & -0.059$\pm$0.082$\pm$0.019 (0.046) \\
n & 0.206 & 0.146$\pm$0.143$\pm$0.041 (0.057) & 0.036$\pm$0.114$\pm$0.032 (0.048) & -0.108$\pm$0.138$\pm$0.039 (0.068) & 0.080$\pm$0.114$\pm$0.033 (0.064) \\
n & 0.265 & -0.050$\pm$0.097$\pm$0.027 (0.034) & 0.039$\pm$0.067$\pm$0.019 (0.028) & -0.129$\pm$0.096$\pm$0.028 (0.049) & 0.059$\pm$0.067$\pm$0.019 (0.037) \\
n & 0.349 & -0.348$\pm$0.153$\pm$0.051 (0.109) & 0.015$\pm$0.076$\pm$0.021 (0.032) & -0.163$\pm$0.133$\pm$0.039 (0.076) & 0.087$\pm$0.077$\pm$0.022 (0.044) \\\hline
\end{tabular}
\centering
  \caption{Tabulated results. Format follows ``central value'' $\pm$ ``statistical uncertainty'' $\pm$ ``experimental systematic uncertainty'' (``model systematic uncertainties'').}
\label{table:results}
\end{table*}

\bibliography{jlab_e06010_prl_sub}

\end{document}